\documentclass[pra,groupedaddress,showpacs,twocolumn,nofootinbib]{revtex4-1}
\usepackage{amsmath,amsfonts,amssymb,amsthm}
\usepackage{graphicx}
\usepackage{epstopdf}

\newtheorem{Def1}{Definition}
\newtheorem{Theorem1}{Theorem}
\newtheorem{Theorem2}[Theorem1]{Theorem}
\newtheorem{Theorem3}[Theorem1]{Theorem}
\newtheorem{Theorem4}[Theorem1]{Theorem}
\newtheorem{Lemma1}{Lemma}
\newtheorem{Lemma2}[Lemma1]{Lemma}
\newtheorem{Lemma3}[Lemma1]{Lemma}

\def\AC{{\cal A}}
\def\BC{{\cal B}}
\def\CC{{\cal C}}
\def\EC{{\cal E}}
\def\GC{{\cal G}}
\def\HC{{\cal H}}
\def\IC{{\cal I}}

\def\TC{{\cal T}}
\def\UC{{\cal U}}

\def\TR{{\bar R}}
\def\bra#1{\langle#1|}
\def\dyad#1#2{|#1\rangle\langle#2|}

\def\ket#1{|#1\rangle }
\def\half{\frac{1}{2}}

\newcommand{\ot}{\otimes }

\newcommand*{\tr}{\mathop{\mathrm{tr}}\nolimits}

\long\def\ca#1\cb{}

 \def\outl#1{}  \def\xa{} \def\xb{}

\ca
 \def\outl#1{\par{\medskip\noindent\hspace*{.5cm}\bf
      \mathversion{bold}#1\mathversion{normal}\smallskip} }
 \long\def\xa#1\xb{}
 
\cb

\ca
 \def\outl#1{\par{\medskip\noindent\hspace*{.5cm}\bf
      \mathversion{bold}#1\mathversion{normal}\smallskip} }
 \def\xa{} \def\xb{}  
\cb

\begin{document}


\title{Fast controlled unitary protocols using group or quasigroup structures}

\author{Li Yu}
\email{cqtliyu@nus.edu.sg}
\affiliation{Department of Physics, Carnegie-Mellon University, Pittsburgh, Pennsylvania 15213, USA}
\affiliation{Centre for Quantum Technologies, National University of Singapore, 3 Science Drive 2, Singapore 117543}

\begin{abstract}
A nonlocal bipartite unitary gate can sometimes be implemented using prior entanglement and only one round of classical communication in which the two parties send messages to each other simultaneously. This cuts the classical communication time by a half compared to the usual protocols, which require back-and-forth classical communication. We introduce such a ``fast'' protocol that can implement a class of controlled unitaries exactly, where the controlled operators form a subset of a projective representation of a finite group, which may be Abelian or non-Abelian. The entanglement cost is only related to the size of the group and is independent of the dimension of the systems. We also introduce a second fast protocol that can implement any given controlled unitary approximately. This protocol uses the algebraic structure of right quasigroups, which are generalizations of quasigroups, the latter being equivalent to Latin squares.  This second protocol could optionally use shared classical randomness as a resource, in addition to using entanglement.  When compared with other known fast unitary protocols, the entanglement cost of this second protocol is lower for general controlled unitaries except for some rare cases.
\end{abstract}

\maketitle

\allowdisplaybreaks


\section{Introduction}\label{sct1}

Entanglement assisted by classical communication and local quantum operations can be used to carry out bipartite nonlocal unitaries. This has been the subject of various studies \cite{Eisert,ReznikNLU,NLU}; for a more extensive list of papers see Ref.~\cite{NLU}. In the paper \cite{nluf} we considered protocols that require less time in classical communication than the usual protocols. In these \emph{fast} protocols, the classical communications from each party to the other are carried out simultaneously. The ability to implement nonlocal unitaries rapidly may be helpful in reducing the effects of noise and decoherence in distributed quantum computation \cite{Cirac_dist,Yims,Yims2,Meter}. Also, the fast unitary protocols have found applications in position-based quantum cryptography \cite{Kent,Kent06,Lau,Buhrman,Buhrman2,Speelman}, where they are used to attack certain position verification schemes.

Some works in the literature on this topic include Groisman and Reznik \cite{Groisman05} for a CNOT gate on two qubits, and Dang and Fan \cite{Dang} for its counterpart on two qudits. In addition Buhrman {\it et al.}\ \cite{Buhrman} and Beigi and K\"onig \cite{Beigi11} have discussed approximate schemes for what they call ``instantaneous quantum computation'', equivalent to a fast bipartite unitary in our language. In Sec.~\ref{sbct5.1} of this paper we will see that the main protocol in Clark {\it et al.} \cite{Clark10} for nonlocal measurements can also be adapted to a fast unitary protocol.

In \cite{nluf}, we had found fast protocols for the following two types of bipartite unitaries: controlled unitaries of the form shown in \eqref{eqn1} below, but with the restriction that the controlled operators form a subset of an ordinary representation of an Abelian group (or a subset of a projective representation of a cyclic group); and certain unitaries of the form $ {\cal U}=\sum_{f\in G}c(f)\,U(f)\otimes V(f)$ where $G$ is a finite group, $c(f)$ are complex numbers, $U(f)$ and $V(f)$ are unitaries on Alice's and Bob's Hilbert spaces, respectively, and the operators $U(f)\ot V(f)$ form a projective representation of $G$.  The protocols for these two types of unitaries are different from each other, but both require the use of a maximally entangled state of Schmidt rank $N$, where $N$ is the size of the group. We also showed that by increasing the amount of entanglement expended, certain classes of bipartite unitaries (including those diagonal in a product of bases) can be approximately carried out using these fast protocols.

In this paper, we first introduce a more general fast controlled unitary protocol in Sec.~\ref{sct2}, where the controlled operators form a subset of a projective representation of a finite group. This improves upon the result of \cite{nluf} in that the group can be non-Abelian, and the representation can be projective (as opposed to ordinary). We then introduce a modified protocol that uses a right quasigroup structure in Sec.~\ref{sct3}, which is shown in Sec.~\ref{sct4} to be able to implement \emph{any} controlled unitary approximately.

We have studied the entanglement cost of these fast protocols. Our protocols in Secs. \ref{sct2} and \ref{sct3} require the use of a maximally entangled state of Schmidt rank $N$, where $N$ is the size of the group or right quasigroup. For general controlled unitaries, we have derived an expression for the entanglement cost of the approximate fast protocol in Sec.~\ref{sct4}, which is a function of the dimensions of the systems and the allowed error, see Theorem ~\ref{thm3}. This cost is usually smaller than other protocols in the literature, but in the special case of very small allowed error (doubly-exponentially small in the dimensions of the systems), the cost of our protocol is generally higher than that of the adapted Clark {\it et al.} protocol, see Sec.~\ref{sct5}.

The paper is organized as follows. As mentioned above, Sections \ref{sct2} and \ref{sct3} include the two fast protocols.  Section~\ref{sbct4.1} applies the fast protocol in Sec.~\ref{sct3} to approximately implement \emph{any} controlled unitary, and calculate the entanglement cost. Section~\ref{sbct4.2} includes an example showing that the entanglement cost can be reduced for some special cases.  In Sec.~\ref{sct5} we first describe how the instantaneous nonlocal measurement protocol in \cite{Clark10} can be adapted to a fast unitary protocol, and then compare the entanglement cost of various fast unitary protocols when they are used to implement controlled unitaries. The concluding Sec.~\ref{sct6} contains a brief summary along with some open problems.

\section{Fast protocol for controlled-group unitaries}\label{sct2}

In this section we construct a fast protocol for any controlled unitary of the form
\begin{equation}\label{eqn1}
\UC=\sum_{k\in S} P_k \otimes V_k,
\end{equation}
where the $P_k$ are orthogonal projectors, possibly of rank greater than 1, on
a Hilbert space $\HC_A$ of dimension $d_A$; $S$ is a subset of a finite group $G$,
and the $\{V_k:k\in S\}$ are unitary operators on a Hilbert space $\HC_B$ of dimension $d_B$ that form a
subset of a projective representation of a group $G$ of order $N$.  As shown in
Appendix~A2 of \cite{nluf}, it suffices to consider the case that the projectors $P_k$ are of rank 1. That is, a scheme for implementing
\begin{equation}\label{eqn2}
\UC=\sum_{k\in S} \dyad{k}{k} \otimes V_k,
\end{equation}
where $\ket{k}$ denotes a ket belonging to a standard (or computational)
orthonormal basis, is easily extended to one that carries out the more general
unitary in \eqref{eqn1} with higher rank $P_k$.  Here is a rough summary of such an extension (for details see Appendix~A2 of \cite{nluf}): first, do a local unitary on system $A$ and an ancillary system $E$,
to transfer the information about ``which projector'' from $A$ to $E$, then implement the protocol for controlled unitaries of the form \eqref{eqn2} (with rank-one projectors on $E$) on systems $E$ and $B$,
and finally do the inverse of the initial local unitary to send back the information about which projector to $A$.  In what follows, we will show how to implement unitaries of the form \eqref{eqn2} using simultaneous classical communication in both directions.

\begin{figure}[ht]
\begin{centering}
\includegraphics[bb=144 686 390 800, width=9.0 cm, clip]{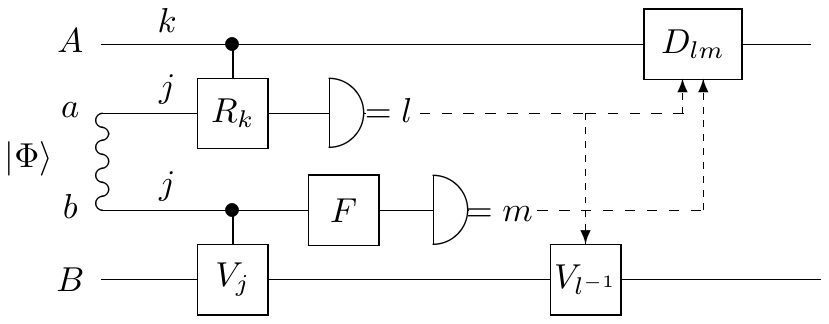}
\caption{The fast protocol for implementing the unitary $\UC=\sum_{k\in S} \dyad{k}{k} \otimes V_k$, where $\{V_k:k\in S\}$ is a subset of a projective representation $\{V_j:j\in G\}$ of a finite group $G$ of order $N$, and the controlled-$V_j$ gate in the circuit involves all $j\in G$.}
\label{fgr1}
\end{centering}
\end{figure}

\subsection{The case of $S=G$}\label{sbct2.1}

Let us first consider the case that $\{V_k: k\in S\}$ in \eqref{eqn2} are a projective representation of a group $G$, i.e. the case $S=G$. (For some background knowledge about projective representations, see Chap.~12 of \cite{GroupBook}.) Each integer $k$ between 0 and $N-1$ identifies an element of the group, and we denote by $e$ the integer labeling the identity element of the group. Assume the factor system of the projective representation is $\{\lambda(g,h)\}$, defined through the following equation:
\begin{equation}\label{eqn3}
 V_g V_h = \lambda(g,h) V_{g*h},
\end{equation}
where $g*h$ denotes the group product of the group elements $g$ and $h$.
As $e$ labels the identity element of the group, $V_e$ is proportional to the identity matrix. And without loss of generality we assume $V_e$ is exactly equal to the identity matrix. The protocol is shown in Fig.~\ref{fgr1} (the figure is for both Secs.~\ref{sbct2.1} and \ref{sbct2.2}), where
\begin{equation}\label{eqn4}
\ket{\Phi} = \frac{1}{\sqrt{N}}\sum_{j=0}^{N-1} \ket{j}_a\otimes \ket{j}_b
\end{equation}
is a fully entangled state on the ancillary systems $a$ and $b$ associated
with $A$ and $B$, respectively, hence the entanglement cost of this protocol is $\log_2 \vert G\vert=\log_2 N$ ebits. The gates $R_k$, $F$ and $D_{lm}$ are defined by
\begin{align}\label{eqn5}
R_k &= \sum_{j=0}^{N-1} \frac{\lambda(k,j^{-1})}{\lambda(j^{-1},j)}\dyad{j* k^{-1}}{j},\notag\\
F &= \frac{1}{\sqrt{N}}\sum_{m,j=0}^{N-1}e^{2\pi i m j/N}\dyad{m}{j}, \notag\\
D_{lm} &= \sum_{k=0}^{N-1} e^{-2\pi i m (l * k) /N} \dyad{k}{k},
\end{align}
where $j^{-1}$ is the group inverse of $j$, and is labeled by an integer between $0$ and $N-1$, like any other element of the group; the multiplication between $m$ and $j$ or between $m$ and $(l*k)$ is the usual multiplication of integers.

The protocol proceeds as follows: Alice carries out a controlled-$R_k$ gate on her systems $A$ and $a$, and Bob carries out a controlled-$V_k$ gate on his systems $b$ and $B$. Then Alice measures $a$ in the standard basis, with the outcome denoted by $l$ ($0\le l \le N-1$), and at the same time Bob does a Fourier gate (denoted by $F$) on $b$ and measures in the standard basis, with the outcome denoted by $m$ ($0\le m \le N-1$). They each send the local measurement outcome to the other party, and then Alice does a unitary $D_{lm}$ gate on $A$, and Bob does a unitary $V_{l^{-1}}$ gate on $B$. They have then implemented $\UC$ on $AB$.

Note that there are other choices for the gate $F$ (when $F$ changes, the $D_{lm}$ gate changes accordingly), and we have chosen the Fourier gate for simplicity. In the controlled-Abelian-group protocol in our previous paper \cite{nluf}, we had used some other gate on $b$ in place of the Fourier gate, and accordingly the final local correction on $A$ was different from the $D_{lm}$ above. But since our current protocol works for any group $G$, it will also work for the special case of Abelian groups, hence the current choice of $F$ and $D_{lm}$ represent another way of implementing the controlled-Abelian-group unitaries.

Now we show a calculation about how the input state evolves under the protocol. The input state on $AB$ can be written in the form
\begin{equation}\label{eqn6}
\ket{\Psi}_{AB}=\sum_{k\in S} \ket{k}_A \otimes \ket{\psi_k}_B,
\end{equation}
where $S\subset G$. We assume $\ket{\Psi}_{AB}$ is normalized, then since $\ket{k}_A$ are normalized, the kets $\ket{\psi_k}$ are generally not normalized. (Although $S=G$ for the current case of a whole representation, we deliberately use $S$ instead of $G$ to illustrate that the protocol still works in the case of a subset of a representation discussed in Sec.~\ref{sbct2.2}.)
The state evolves under the protocol as follows:
\begin{align}\label{eqn7}
&\frac{1}{\sqrt{N}}\sum_{j=0}^{N-1}\sum_{k\in S} \ket{j}_a\otimes \ket{j}_b \otimes \ket{k}_A \otimes \ket{\psi_k}_B \notag\\
&\xrightarrow{R_k,\,V_j} \frac{1}{\sqrt{N}}\sum_{j=0}^{N-1}\sum_{k\in S} \frac{\lambda(k,j^{-1})}{\lambda(j^{-1},j)} \notag\\
& \quad\quad\quad\ket{j* k^{-1}}_a\otimes \ket{j}_b \otimes \ket{k}_A \otimes V_j\ket{\psi_k}_B \notag\\
&\xrightarrow{F} \frac{1}{N}\sum_{j=0}^{N-1}\sum_{m=0}^{N-1}\sum_{k\in S} e^{2\pi i m j/N} \frac{\lambda(k,j^{-1})}{\lambda(j^{-1},j)} \notag\\
& \quad\quad\quad\ket{j* k^{-1}}_a\otimes \ket{m}_b \otimes \ket{k}_A \otimes V_j\ket{\psi_k}_B \notag\\
&\xrightarrow[l=j* k^{-1}]{\mbox{measure}\,\,a,b} \sum_{k\in S} e^{2\pi i m (l* k)/N} \frac{\lambda(k,k^{-1}* l^{-1})}{\lambda(k^{-1}* l^{-1},l* k)} \notag\\
& \quad\quad\quad \ket{k}_A \otimes V_{l* k}\ket{\psi_k}_B \notag\\
&\xrightarrow{D_{lm}} \sum_{k\in S}
\frac{\lambda(k,k^{-1}* l^{-1})}{\lambda(k^{-1}* l^{-1},l* k)}
\ket{k}_A \otimes V_{l* k}\ket{\psi_k}_B \notag\\
&\xrightarrow[\mbox{let}\,\,j:=l* k]{V_{l^{-1}}} \sum_{k\in S}
\frac{\lambda(k,j^{-1})}{\lambda(j^{-1},j)}
\ket{k}_A \otimes V_{k* j^{-1}}V_j\ket{\psi_k}_B \notag\\
&=\sum_{k\in S}\frac{1}{\lambda(j^{-1},j)}
\ket{k}_A \otimes V_k V_{j^{-1}}V_j\ket{\psi_k}_B \notag\\
&=\sum_{k\in S}\ket{k}_A \otimes V_k V_e\ket{\psi_k}_B \notag\\
&=\sum_{k\in S} \ket{k}_A \otimes V_k\ket{\psi_k}_B=\UC \ket{\Psi}_{AB},
\end{align}
where in deriving the second-to-last and third-to-last lines we have used the defining equation \eqref{eqn3} for factors in a factor system. In deriving the fourth line (measurement of $a,b$), we have used a normalization factor $N$ which is independent of the measurement outcomes $l,m$. This factor is chosen to be $N$ in order for the resulting state to be normalized (to see that the resulting state is normalized, note that it is in the form of a controlled unitary acting on the original input state $\sum_{k\in S} \ket{k}_A \otimes \ket{\psi_k}_B$).  This choice of normalization factor means that each pair of the measurement outcomes $(l,m)$ occurs with the same probability $1/N^2$, hence the marginal probability of each $l$ occurring is $1/N$. This last result is helpful in analyzing the modified protocol in Sec.~\ref{sct3}, see the remarks after Eq.~\eqref{eqn18}.\\

\textbf{Example 1.} Suppose $d_B=2$ and $\UC=\sum_{k=0}^{3} P_k \otimes \sigma_k$, where $P_k$ are orthogonal projectors on system $A$, and $\sigma_k$ are the four Pauli operators on system $B$. Take $G$ to be the $C_2\times C_2$ group of order 4 (the Klein four-group). We can implement $\UC$ by implementing $\UC'=\sum_{k=0}^{3} \dyad{k}{k}_E \otimes \sigma_k$ on system $EB$ using the protocol above and doing local unitaries on $AE$ before and after $\UC'$, see the remarks after Eq.~\eqref{eqn2}.  The resource state needed is a maximally entangled state of Schmidt rank 4.

\subsection{Subset of a group representation}
\label{sbct2.2}

Assume that the $\{V_k\}$ form a projective representation of a group of order $N$,
but the sum over $k$ in \eqref{eqn1} is restricted to some subset $S$ of the set of integers between $0$ and $N-1$. It will suffice once again to consider the case of rank-one projectors, i.e., \eqref{eqn2}. The protocol has the same circuit diagram as shown in Fig.~\ref{fgr1}, but with $R_k$'s restricted to those $k\in S$. The state evolution calculated in \eqref{eqn7} still holds, hence the protocol still works. The entanglement cost is still $\log_2 \vert G\vert=\log_2 N$ ebits. This discussion of the ``subset'' case is similar to that in Sec.~II~C of \cite{nluf}, where the case of a subset of an Abelian group representation was discussed.\\

\textbf{Example 2.}
Similar to Example 1, let $V_k$ be the Pauli operators, but now not all of them need to appear in $\UC$. One such bipartite unitary is $\UC=P_0\otimes I_B + P_1\otimes X_B+P_2\otimes Z_B$. The group $G$ is still the $C_2\times C_2$ group, and the entanglement cost of our protocol is still 2 ebits.

\section{Fast protocol using a right quasigroup}
\label{sct3}

\begin{figure}[ht]
\begin{centering}
  \includegraphics[bb=144 686 390 800, width=9.0 cm, clip]{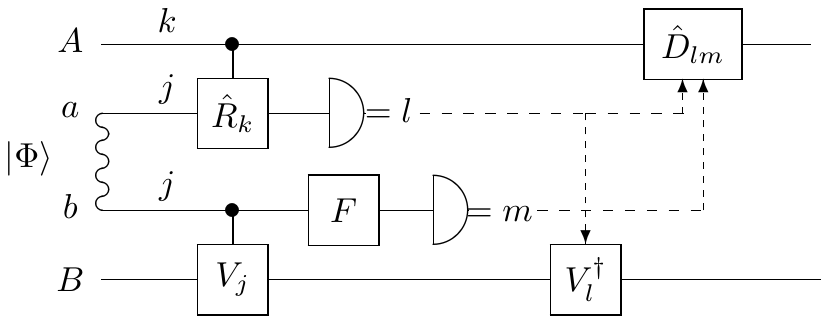}
\end{centering}
\caption{The fast protocol for approximately implementing the unitary
$\UC=\sum_{k\in S} \dyad{k}{k} \otimes V_k$, where $S$ is a subset of a finite right quasigroup $Q$, and $\{V_k: k\in Q\}$ is an approximate unitary representation of $Q$ (see Definition \ref{defi1}).
The resource entangled state is $\ket{\Phi} = \frac{1}{\sqrt{N}}\sum_{j=0}^{N-1} \ket{j}_a\otimes \ket{j}_b$. The protocol actually implements a unitary $\UC_l$ dependent on the measurement outcome $l$ on system $a$.}
\label{fgr2}
\end{figure}

\begin{figure}[ht]
\begin{centering}
  \includegraphics[bb=144 630 416 780, width=9.0 cm, clip]{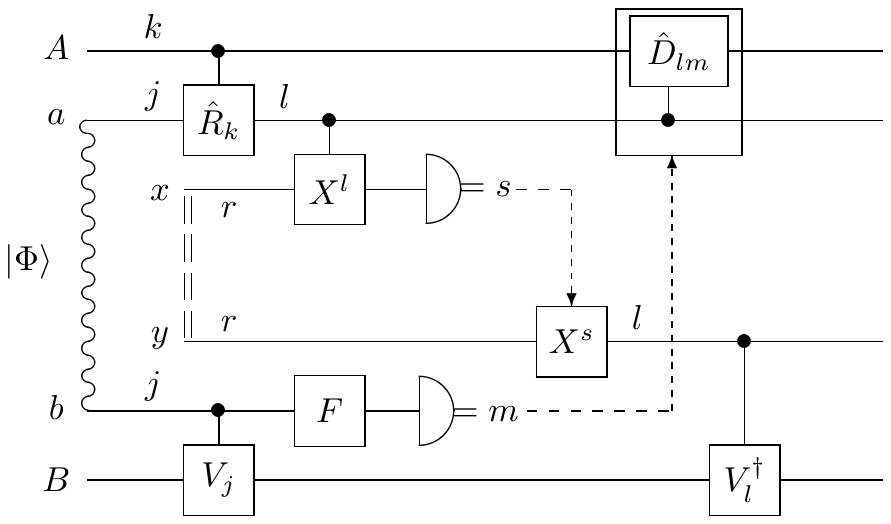}
\end{centering}
\caption{A modified version of the protocol shown in Fig.~\ref{fgr2} that hides the value of $l$.  The systems $x,y$ are initially in the mixed state $\frac{1}{N}\sum_{r=0}^{N-1} \dyad{r}{r}_x \otimes \dyad{r}{r}_y$. The $X$ gate in the circuit is $X=\sum_{j=0}^{N-1} \dyad{(j-1)\mbox{ mod }N}{j}$. The quantum controlled-$\hat D_{lm}$ gate (where system $a$ is controlling using the standard basis states $\ket{l}$) is dependent on the classical message $m$. The systems $a$ and $y$ are not measured at the end.}
\label{fgr3}
\end{figure}

The use of groups and their unitary representations proved to be very helpful in designing various protocols for nonlocal unitaries (e.g. \cite{NLU,nluf} and the protocol in the previous section). In this section we introduce two more general types of algebraic structures named \emph{quasigroup} and \emph{right quasigroup}, and define an ``approximate unitary representation'' of right quasigroups, and then describe a fast approximate controlled unitary protocol (with two versions shown in Figs.~\ref{fgr2} and \ref{fgr3}) where the controlled operators form a subset of such a representation. After that, we analyze the error in the protocol. Finally we discuss an extended case where some controlled operators are the same.

A quasigroup is an algebraic structure resembling a group, but the multiplication need not be associative and there need not be an identity element. More explicitly (see \cite{quasi}), a quasigroup ($Q$, *) is a set $Q$ with a binary operation *, with the property that for any two elements $a$ and $b$ in $Q$, there exist unique elements $x$ and $y$ in $Q$ such that
\begin{align}
 &a * x = b, \label{eqn8}\\
 &y * a = b. \label{eqn9}
\end{align}
The multiplication table (Cayley table) of a quasigroup is a Latin square, the latter defined as an $n\times n$ array filled with $n$ different symbols, each occurring exactly once in each row and exactly once in each column. And each Latin square is the multiplication table of some quasigroup. A special type of quasigroup is called the \emph{loop}. A loop is a quasigroup with an identity element $e$ such that
\begin{equation}\label{eqn10}
 x*e = x = e*x,\,\,\forall x\in Q.
\end{equation}
It follows that the identity element $e$ is unique, and that every element of $Q$ has a unique left inverse and a unique right inverse. A \emph{right quasigroup} ($Q$, *) is a set $Q$ with a binary operation * that only satisfies \eqref{eqn9} and does not need to satisfy \eqref{eqn8}. A right quasigroup with an identity element satisfying \eqref{eqn10} is called a \emph{right loop}.

Next we introduce the concept of an ``approximate unitary representation'' of a right quasigroup which is useful for the nonlocal unitary protocols.

\begin{Def1}\label{defi1}
(Approximate unitary representation of a right quasigroup) Define the $(\eta,\delta)$-approximate unitary representation of a right quasigroup $(Q,*)$ to be a set of unitary operators $\{V_i: i\in Q\}$ satisfying that for any $k\in Q$,
\begin{align}\label{eqn11}
& \| V_{l(j,k)} V_k - V_j\|_\infty <\eta,\, \notag\\
& \mbox{ for at least } N(1-\delta) \mbox{ distinct values of } j,
\end{align}
where $N$ is the size of $Q$, $l(j,k)$ is the unique element in $Q$ that satisfies $l(j,k)*k=j$, and $\eta$ and $\delta$ are positive constants. The $\| \cdot \|_\infty$ notation in \eqref{eqn11} denotes the maximum singular value of an operator.
\end{Def1}

Note that the $\| \cdot \|_\infty$ norm in this paper is the Schatten p-norm with $p=\infty$. (This should not be confused with the induced norm, see \cite{norms} for the definition of both types of norms; in fact our norm is the same as the induced norm with subscript 2, instead of $\infty$.)

Now we introduce a fast protocol, shown in Fig.~\ref{fgr2}, that approximately implements controlled unitaries of the form
\begin{equation}\label{eqn12}
\UC=\sum_{k\in S} \dyad{k}{k} \otimes V_k,
\end{equation}
where $\{\ket{k}\}$ is an orthonormal basis, and the controlled operators $V_k$ form a subset $S$ of an $(\eta,\delta)$-approximate unitary representation of a finite right quasigroup $Q$ of size $N$. The other matrices in the representation are also denoted by $V_k$, for $k\in Q\backslash S$. The whole representation is denoted by $\{V_k:k\in Q\}$. In Eq.~\eqref{eqn12}, every element of the right quasigroup corresponds to at most one term in the expansion, but the protocol can be extended to the more general case where some terms correspond to the same quasigroup element, see the last paragraph of this section. Both $\eta$ and $\delta$ in \eqref{eqn11} are related to the closeness of the implemented quantum operation to the desired unitary, see \eqref{eqn32}. The circuit diagram for the protocol is shown in Fig.~\ref{fgr2}.

In Fig.~\ref{fgr2}, the entangled state $\ket{\Phi}$ is still given by Eq.~\eqref{eqn4}, hence the entanglement of this protocol is $\log_2 \vert Q\vert=\log_2 N$ ebits. In order to define the gates in the circuit, we label each element of $Q$ using a unique integer between $0$ and $N-1$. Any such labeling scheme will work.  The $F$ and controlled-$V_j$ gates are the same as those in Fig.~\ref{fgr1}. Since $V_l$ is unitary, $V_l^\dag$ is equal to $(V_l)^{-1}$. The $\hat R_k$ and $\hat D_{lm}$ gates are defined as follows:
\begin{align}\label{eqn13}
&\hat R_k = \sum_{j\in Q} \dyad{l(j,k)}{j},\notag\\
&\hat D_{lm} = \sum_{k\in S} e^{-2\pi i m (l*k) /N} \dyad{k}{k},
\end{align}
where $l$ and $m$ are measurement outcomes on $a$ and $b$, respectively; $l$ is equal to $l(j,k)$, the unique element in $Q$ that satisfies $l(j,k)*k=j$. In the second equation in \eqref{eqn13}, $l*k$ is the product of $l$ and $k$ in $Q$; the multiplication between $m$ and $(l*k)$ is the usual multiplication of integers. From the definition of a right quasigroup, $j=l*k$ is unique for fixed $l$ and $k$, and since $l$ is uniquely determined by $j$ and $k$, each $\hat R_k$ is a permutation matrix.

Hereafter we assume $S=\{0,1,\cdots,M-1\}$ for some $M\le N$, which can be obtained by relabeling the group elements $k$, and it is safe because there is no requirement that $k=0$ denotes the identity element in $Q$. With the input state on $AB$ defined as $\ket{\Psi}_{AB}:=\sum_{k=0}^{M-1} \ket{k}_A \otimes \ket{\psi_k}_B$, the same as in Eq.~\eqref{eqn6}, the state evolution of this protocol is as follows:
\begin{align}\label{eqn14}
&\frac{1}{\sqrt{N}}\sum_{j=0}^{N-1}\sum_{k=0}^{M-1} \ket{j}_a\otimes \ket{j}_b \otimes \ket{k}_A \otimes \ket{\psi_k}_B \notag\\
&\xrightarrow{\hat R_k,\,V_j} \frac{1}{\sqrt{N}}\sum_{j=0}^{N-1}\sum_{k=0}^{M-1} \ket{l(j,k)}_a\otimes \ket{j}_b \otimes \ket{k}_A \otimes V_j\ket{\psi_k}_B \notag\\
&\xrightarrow{F} \frac{1}{N}\sum_{j=0}^{N-1}\sum_{m=0}^{N-1}\sum_{k=0}^{M-1}
e^{2\pi i m j/N} \notag\\
&\quad\quad \ket{l(j,k)}_a\otimes \ket{m}_b \otimes \ket{k}_A \otimes V_j\ket{\psi_k}_B \notag\\
&\xrightarrow{\mbox{measure}\,\,a,b} \sum_{k=0}^{M-1} e^{2\pi i m (l*k)/N}
\ket{k}_A \otimes V_{(l*k)}\ket{\psi_k}_B \notag\\
&\xrightarrow{\hat D_{lm}} \sum_{k=0}^{M-1}
\ket{k}_A \otimes V_{(l*k)}\ket{\psi_k}_B \notag\\
&\xrightarrow{V_l^\dag} \sum_{k=0}^{M-1}
\ket{k}_A \otimes V_l^\dag V_{(l*k)}\ket{\psi_k}_B \notag\\
&=\sum_{k=0}^{M-1} \ket{k}_A \otimes (V_k+E_{k,l})\ket{\psi_k}_B=\UC_l \ket{\Psi}_{AB},
\end{align}
where in the last line
\begin{align}\label{eqn15}
E_{k,l} := V_l^\dag V_{(l*k)}-V_k, \,\,\forall k\in S, l\in Q,\\
\UC_l:=\sum_{k=0}^{M-1}\dyad{k}{k}_A \otimes V_l^\dag V_{(l*k)}, \,\,\forall l\in Q. \label{eqn16}
\end{align}

From \eqref{eqn11}, and using $\|V_l^\dag (V_{(l*k)} - V_l V_k )\|_\infty=\|V_{(l*k)} - V_l V_k\|_\infty$, we get that for any fixed $k$,
\begin{align}\label{eqn17}
& \|E_{k,l}\|_\infty<\eta,\, \notag\\
& \mbox{ for at least } N(1-\delta) \mbox{ distinct values of } l.
\end{align}

By the definition of $E_{k,l}$, we also have
\begin{equation}\label{eqn18}
\| E_{k,l}\|_\infty = \| V_l^\dag V_{(l*k)}-V_k\|_\infty \le \| V_l^\dag V_{(l*k)}\|_\infty +\| V_k\|_\infty = 2.
\end{equation}

From Eq.~\eqref{eqn16} and the last line of Eq.~\eqref{eqn14}, we see that the protocol shown in Fig.~\ref{fgr2} implements one unitary in the set of unitaries $\{\UC_l\}$ depending on the measurement outcome $l$ (while independent of the outcome $m$ on system $b$).  The probability that a particular outcome $l$ occurs is $\frac{1}{N}$. This probability is the same for all $l$, because all $\hat R_k$ are permutation matrices. A more technical explanation is obtained by looking at the remarks after Eq.~\eqref{eqn7} that each $l$ occurs with probability $1/N$, and noting that a similar argument also holds for Eq.~\eqref{eqn14}. Therefore, the average quantum operation performed on $AB$ is
\begin{align}\label{eqn19}
\EC(\rho_{AB})&=\frac{1}{N}\sum_{l=0}^{N-1} \UC_l \rho_{AB}\UC_l^\dag \notag\\
&=\frac{1}{N}\sum_{l=0}^{N-1} \sum_{k,k'=0}^{M-1} \{[\dyad{k}{k}\otimes (V_k+E_{k,l})] \notag\\
& \rho_{AB} [\dyad{k'}{k'}\otimes (V_{k'}+E_{k',l})^\dag]\}.
\end{align}

Next, define our ideal quantum operation as follows:
\begin{equation}\label{eqn20}
\EC_\UC(\rho_{AB}):=\UC \rho_{AB} \UC^\dag.
\end{equation}
In the following we show that the superoperators $\EC$ and $\EC_\UC$ are arbitrarily close to each other when $\eta$ and $\delta$ approach 0, although some $\UC_l$ might not be close to $\UC$. The distance between $\EC$ and $\EC_\UC$ will be measured using the diamond norm of their difference: $\|\EC-\EC_\UC\|_\diamond$, where the diamond norm is defined in Eq.~\eqref{eqn23} below, see also Eq.~(7) of \cite{Beigi11}. We firstly define the trace norm of an operator as follows:
\begin{equation}\label{eqn21}
\|L\|_1 = \tr\sqrt{L^\dagger L}\ ,\qquad L\in \BC(T_1)\ .
\end{equation}
where $\BC(T_1)$ is the space of bounded operators on a complex Hilbert space $T_1$. The trace norm induces a norm
\begin{equation}\label{eqn22}
\|\Omega\|_1 =\max_{L\in\BC(T_1): \|L\|_1\leq 1} \|\Omega(L)\|_1
\end{equation}
on the set of superoperators $\Omega:\BC(T_1)\rightarrow\BC(T_2)$, where $T_2$ is a complex Hilbert space that may be different from $T_1$. The diamond norm is defined as
\begin{equation}\label{eqn23}
\|\Omega\|_\diamond =\sup_{k\geq 1}\|\Omega\otimes\IC_{\mathbb{C}^k}\|_1\
\end{equation}
where $\IC_{\mathbb{C}^k}$ is the identity superoperator on $\BC(\mathbb{C}^k)$, where $\mathbb{C}^k$ is a $k$-dimensional complex Hilbert space.

Theorem 6 in \cite{Kret08} provides a general upper bound for the diamond norm of the difference of two superoperators. Setting the $\EC$ and $\EC'$ in that theorem to be the $\EC_\UC$ and $\EC$ in this paper, respectively, we get that
\begin{equation}\label{eqn24}
\|\EC_\UC-\EC\|_\diamond \le 2\| U'-V'\|_\infty,
\end{equation}
where $U'$ is an isometric dilation for $\EC_\UC$, and $V'$ is an isometric dilation for the quantum operation $\EC$ in this paper:
\begin{align}\label{eqn25}
U' &= \frac{1}{\sqrt{N}}\sum_{l=0}^{N-1}\UC\otimes \dyad{l}{0},\notag\\
V' &= \frac{1}{\sqrt{N}}\sum_{l=0}^{N-1}\UC_l \otimes \dyad{l}{0}.
\end{align}
In \eqref{eqn25}, the $\UC$ and $\UC_l$ act on the combined $AB$ system, and the operators $\dyad{l}{0}$ act on an ancillary system $R$. The set of kets $\{\ket{l}\}$ is (a part of) an orthonormal basis. Then
\begin{align}\label{eqn26}
& \| U'-V'\|_\infty \notag\\
&= \| \frac{1}{\sqrt{N}}\sum_{l=0}^{N-1}(\UC-\UC_l)\otimes \dyad{l}{0} \|_\infty \notag\\
&= \| \frac{1}{\sqrt{N}}\sum_{l=0}^{N-1}\sum_{k=0}^{M-1}\dyad{k}{k}\otimes (-E_{k,l})\otimes \dyad{l}{0} \|_\infty \end{align}
Denote the operator inside the norm symbol in the last line by $K$. Since $\| K \|_\infty$ is the square root of the maximum eigenvalue of $K^\dag K$, in the following we will consider the operator
\begin{equation}\label{eqn27}
K^\dag K = \frac{1}{N} \sum_{k=0}^{M-1}\sum_{l=0}^{N-1} \dyad{k}{k}\otimes E_{k,l}^\dag E_{k,l}\otimes \dyad{0}{0}.
\end{equation}
Let $\lambda_0(A)$ denote the maximum eigenvalue of a Hermitian operator $A$, then
\begin{align}\label{eqn28}
& \lambda_0(K^\dag K) \notag\\
& =\lambda_0(\frac{1}{N} \sum_{k=0}^{M-1}\sum_{l=0}^{N-1} \dyad{k}{k}\otimes E_{k,l}^\dag E_{k,l}) \notag\\
& \le  \frac{1}{N} \max_{0\le k\le M-1}\sum_{l=0}^{N-1} \lambda_0(\dyad{k}{k}\otimes E_{k,l}^\dag E_{k,l}) \notag\\
& =  \frac{1}{N} \max_{0\le k\le M-1}\sum_{l=0}^{N-1} \lambda_0(E_{k,l}^\dag E_{k,l}) \notag\\
& = \frac{1}{N} \max_{0\le k\le M-1} [\sum_{(k,l)\in \CC} \lambda_0(E_{k,l}^\dag E_{k,l}) + \sum_{(k,l)\notin \CC} \lambda_0(E_{k,l}^\dag E_{k,l})] \notag\\
& \le \frac{1}{N} \max_{0\le k\le M-1} [N(1-\delta) \cdot \eta^2 + N\delta \cdot 2^2] \notag\\
& \le \eta^2 + 4\delta
\end{align}
where in deriving the third line we have used the fact that the eigenvalues of a block diagonal matrix (where each block is a square matrix) are just the eigenvalues of the individual blocks. In the fifth line the $\CC$ is the set of $(k,l)$ pairs that satisfy $\| E_{k,l}\|_\infty < \eta$, see \eqref{eqn17}. The inequality in the second-last line follows from that $\|E_{k,l}\|_\infty^2$ is the maximum eigenvalue of $E_{k,l}^\dag E_{k,l}$, and also \eqref{eqn17} and the inequality $\| E_{k,l}\|_\infty \le 2$ in \eqref{eqn18}.  It follows from Eq.~\eqref{eqn28} that
\begin{equation}\label{eqn29}
\| K \|_\infty=\sqrt{\lambda_0(K^\dag K)} \le \sqrt{\eta^2 + 4\delta},
\end{equation}
and then using Eqs.~\eqref{eqn24} and \eqref{eqn26} we get that for the protocol shown in Fig.~\ref{fgr2},
\begin{equation}\label{eqn30}
\|\EC_\UC-\EC\|_\diamond \le 2\| U'-V'\|_\infty \le 2 \sqrt{\eta^2 + 4\delta}.
\end{equation}

As mentioned previously, the protocol shown in Fig.~\ref{fgr2} deviates from our original goal in that it implements $\UC_l$ depending on the measurement outcome $l$; and for any given $l$, the operators $\UC_l$ and $\UC$ might not be close to each other. \footnote{The distance between $\UC$ and $\UC_l$ depends on the measure we use: $\|\UC-\UC_l\|_\infty$ can be some large constant (say $>1/2$) for all $l$, but for some other measures of the distance between $\UC$ and $\UC_l$, such as $\vert \bra{\psi}\UC^\dag \UC_l \ket{\psi}\vert^2$ averaged over Haar random $\ket{\psi}$, large distances can only occur for a small fraction of the possible values of $l$.}  To resolve this problem, one method is to assume the protocol is run by  automated machines and the value of $l$ is erased from the memory of the machines at both parties in the end, hence the value of $l$ is unknown to the experimentalist.  The other method is to change the protocol to remove the explicit measurement of $l$. This is illustrated by the modified protocol shown in Fig.~\ref{fgr3}.  The ancillary systems $x$ and $y$ are initialized in a mixed state $\frac{1}{N}\sum_{r=0}^{N-1} \dyad{r}{r}_x \otimes \dyad{r}{r}_y$ which represents shared classical randomness.  The $X$ gate in the circuit is $X=\sum_{j=0}^{N-1} \dyad{(j-1)\mbox{ mod }N}{j}$.  The operations in Fig.~\ref{fgr3} hide the value of $l$ throughout the protocol, and in the end the $a$ and $y$ systems are thrown away, hence there is no information about $l$ that is revealed.  The protocol in Fig.~\ref{fgr3} has the same gates on $AB$ as those shown in Fig.~\ref{fgr2}, and the key difference in the two circuits is on the added operations that involve systems $x$ and $y$, which help carry out $V_l^\dag$ on system $B$ without revealing the value of $l$.  The protocol would implement the desired quantum operation on $AB$ shown in Eq.~\eqref{eqn19} if a hypothetical $Z$-basis measurement on system $y$ just prior to the controlled-$V_l^\dag$ gate on system $yB$ yields the outcome $l$ with probability $1/N$, for each $l\in\{0,1,\cdots,N-1\}$. To see that this is indeed the case, note that according to the remarks after Eq.~\eqref{eqn18}, the probability of getting $l$ on a hypothetical $Z$-basis measurement on system $a$ is $1/N$ for all $l\in\{0,1,\cdots,N-1\}$, and for each hypothetical $\ket{l}$ state on system $a$, the following calculation shows that the middle part of the circuit involving system $xy$ is such that the $\ket{l}$ state appears on system $y$ just prior to the controlled-$V_l^\dag$ gate:
\begin{align}\label{eqn31}
&\frac{1}{N} \sum_{r=0}^{N-1} \dyad{r}{r}_x \otimes \dyad{r}{r}_y \notag\\
&\xrightarrow{X^l} \frac{1}{N}\sum_{r=0}^{N-1} \dyad{r-l \mbox{ mod N}}{r-l \mbox{ mod }N}_x \otimes \dyad{r}{r}_y \notag\\
&\xrightarrow[s=r-l \mbox{ mod }N]{\mbox{measure }x}  \dyad{l+s \mbox{ mod }N}{l+s \mbox{ mod }N}_y \notag\\
&\xrightarrow{X^s} \dyad{l}{l}_y.
\end{align}
Therefore the protocol shown in Fig.~\ref{fgr3} implements a quantum operation on $AB$ shown in Eq.~\eqref{eqn19}, without revealing the value of $l$.  As the other details of the circuit are the same as in Fig.~\ref{fgr2}, Eq.~\eqref{eqn30} still holds for the protocol shown in Fig.~\ref{fgr3}, hence we have

\begin{Theorem1}\label{thm1}
For the fast controlled unitary protocol shown in Fig.~\ref{fgr3}, with the controlled operators $V_k$ satisfying Definition \ref{defi1}, the ideal quantum operation $\EC_\UC$ and the implemented quantum operation $\EC$ are related in the following way:
\begin{equation}\label{eqn32}
\|\EC_\UC-\EC\|_\diamond \le 2 \sqrt{\eta^2 + 4\delta}.
\end{equation}
\end{Theorem1}


The form of the unitary $\UC$ in \eqref{eqn12} is such that every element of the right quasigroup correspond to at most one term in the expansion. But our protocol can be extended to work for unitaries with more than one appearance of some $V_k$, which are shown as follows:
\begin{equation}\label{eqn33}
\UC=\sum_{i\in \IC} \dyad{i}{i} \otimes V_{k(i)},
\end{equation}
where $k(i)\in Q$ for each $i$, and the index set $\IC$ is larger in size than the set $S$ composed of all different $k(i)$. We call the terms in \eqref{eqn33} with the same $k(i)$ ``redundant terms'', and this notation will also be used in Sec.~\ref{sbct4.1}. There are two ways to implement this $\UC$ approximately:
(1). Combine the redundant terms, such that there are only terms of the form $P_k \otimes V_k$, with $P_k$ being a projector. Then, according to Appendix~A2 of \cite{nluf}, it suffices to consider implementing another unitary where each $P_k$ is replaced by a projector of rank 1. $\quad$(2). (This second method will be used in Sec.~\ref{sct4} below.) Do not combine the terms, and just do the protocol with the local gates in \eqref{eqn13} modified in the following way: All the $k$ in \eqref{eqn13} are replaced by $k(i)$. It is not hard to see that the error analysis preceding Theorem~\ref{thm1} still works, by noting that the number of values of $l$ such that
$\eta \le \| E_{k,l}\|_\infty \le 2 $ is at most $N\delta$ for each $k=k(i)$, and each $l$ value occurs for a fixed probability $1/N$, hence the total contribution from $(k,l)\notin\CC$ in the fourth line of Eq.~\eqref{eqn28} is bounded in exactly the same way as in the case without redundant terms. Therefore $\UC$ is implemented with an error satisfying \eqref{eqn32}.

\section{Fast approximate protocol for any controlled unitary}
\label{sct4}

We now show that the protocol introduced in Sec.~\ref{sct3} can be applied to approximately implement any bipartite controlled unitary in the fast way.

\subsection{Protocol}
\label{sbct4.1}

In this section we only consider the controlled unitaries with rank-1 projectors on the $A$ side:
\begin{equation}\label{eqn34}
\TC=\sum_{i=0}^{M-1} \dyad{i}{i} \otimes W_i,
\end{equation}
where some $W_i$ operators may be equal or very close to each other. In this way we have included controlled unitaries of the form $\TC=\sum_{k=0}^{M-1} P_i \otimes W_i$, where the projectors $P_i$ may be of rank greater than 1; this is because any such projector can be written as a sum of rank-1 projectors, and thus the resulting expansion of $\TC$ fits into the form of Eq.~\eqref{eqn34}. Since phases on the $W_i$'s can be adjusted by a diagonal unitary gate on $\HC_A$, we assume for convenience that $W_i$ all have determinant 1, i.e. they are all in the special unitary group $SU(d_B)$. The general idea of this section is that the unitaries $W_i$ can be approximated by some other unitaries denoted by $V_{k(i)}$, see Eqs.~\eqref{eqn37} and \eqref{eqn38}. These $V_{k(i)}$ are all in the set $\TR_m$ for some integer $m$, where $\TR_m=\{U \big\arrowvert U\in R_m \mbox{ or }U^{-1}\in R_m\}$, and $R_m$ is the set of unitaries defined in Sec.~IV of \cite{Harrow}, see also Appendix~\ref{appB}. It will later be shown that $\TR_m$ is an approximate unitary representation of a finite right quasigroup, hence our fast protocol in Sec.~\ref{sct3} can be applied to implement the controlled unitary $\TC$ approximately.

There are many alternatives to the sets $\TR_m$ for approximating unitaries in $SU(d_B)$, since there are only two requirements for these sets: first, the elements in the set can approximate any unitary in $SU(d_B)$ with some suitable error; second, each such set forms an approximate unitary representation (with some suitable error) of a right quasigroup of a suitable size. Such sets generally are not comprised of the products of some generating unitaries. We choose to use the family $\TR_m$ because it allows a relatively simple proof of the entanglement cost, and the elements of $\TR_m$ are products of unitaries from a small generating set, making the implementation of the protocol easier.

The following lemma describes how a unitary can be approximated by elements of $\TR_m$. The proof is in Appendix~\ref{appC} and makes use of some derivations in \cite{Harrow}.

\begin{Lemma1}\label{lm1}
Let $d>1$ be an integer, and $0<\zeta<1$. There are positive constants $c_0$, $c_1$ and $c_2$ independent of $d$ and $\zeta$, such that for all integers $m$ satisfying
\begin{equation}\label{eqn35}
m > c_0+ c_1 \log{d}+ c_2 \log\frac{1}{\zeta},
\end{equation}
any unitary $U$ in the special unitary group $SU(d)$ can be approximated by a unitary $V\in \TR_m$ within error $\zeta$:
\begin{equation}\label{eqn36}
\| U-V\|_\infty < \zeta.
\end{equation}
\end{Lemma1}

Note that the logarithm in Eq.~\eqref{eqn35} and later parts of the paper is the natural logarithm. Denote $d:=d_B$. With the preparation above, we do the following for the current controlled unitary problem in order to apply our fast protocol in Sec.~\ref{sct3}: for the given $W_i$'s, all of determinant 1, we fix an error parameter $\zeta$, and find the closest approximation of each $W_i$ in terms of an operator in $\TR_m$, where $m=\lceil c_0+ c_1 \log{d}+ c_2 \log\frac{1}{\zeta}\rceil$, with $c_0$, $c_1$ and $c_2$ being positive constants independent of $d$ and $\zeta$. Denote these approximations to $W_i$ by $V_{k(i)}$, $0\le k(i)\le M'-1$, where $M'\le M$. That is,
\begin{equation}\label{eqn37}
\|W_i - V_{k(i)}\|_\infty < \zeta.
\end{equation}
The appearance of a different subscript $k$ and a new bound $M'$ is because some of the $W_i$ matrices may correspond to the same approximation matrix $V_k$. The choice of the approximation matrices define a map from $i$ to $k$, and this map is denoted by $k=k(i)$.  As a result of the approximations above, we get the following approximate version of $\TC$:
\begin{equation}\label{eqn38}
\UC=\sum_{i=0}^{M-1} \dyad{i}{i} \otimes V_{k(i)}
\end{equation}
This expression for $\UC$ contains ``redundant terms'', a terminology introduced at the end of Sec.~\ref{sct3}.

Next, we can always include other unitaries $V_k$ ($M'+1\le k\le N-1$) such that the set $\{V_k:0\le k\le N-1\}$ is the same as $\TR_m$. And in Appendix~\ref{appD}  we will show
\begin{Lemma2}\label{lm2}
Let $d>1$ be the dimension of the Hilbert space under consideration, and $0<\eta<1$, $0<\delta<1$. There is a positive constant $c_3$ independent of $d$, $\eta$ and $\delta$, such that whenever
\begin{equation}\label{eqn39}
m > c_3 \cdot (d^2 \log\frac{2}{\eta}+\log\frac{1}{\delta}),
\end{equation}
there is a right quasigroup $(Q,*)$ for which $\TR_m$ is an $(\eta,\delta)$-approximate unitary representation.
\end{Lemma2}

Therefore we can apply the fast protocol in Sec.~\ref{sct3} to implement the $\UC$ in Eq.~\eqref{eqn38} approximately. In the case there are redundant terms in Eq.~\eqref{eqn38}, i.e. when $k(i)$ is not a one-to-one map, the protocol in Sec.~\ref{sct3} still works with $k$ in \eqref{eqn13} understood as $k(i)$, and the error analysis below is still valid, for basically the same reason as mentioned at the end of Sec.~\ref{sct3}.
What this protocol actually does is to implement a unitary $\UC_l$ depending on the measurement outcome $l$. The $\UC_l$ may be different from each other, and some of them may not even be close to $\UC$, but the average quantum operation is close to $\UC$ under a reasonable measure, as shown in Theorem~\ref{thm2} below. And since $\UC$ is close to $\TC$, this protocol implements $\TC$ approximately.

Now we analyze the size of the set $\TR_m$, which is directly related to the entanglement cost of this protocol, since in the protocol in Sec.~\ref{sct3} the Schmidt rank of the resource entangled state is equal to the size of the right quasigroup, and the latter is equal to the size of $\TR_m$ in our construction. To guarantee that the protocol works with the desired precision, it suffices if the following two conditions hold: $\TR_m$ forms an $(\eta,\delta)$-approximate unitary representation of a right quasigroup; and that each $W_i$ has a good approximation in $\TR_m$ with error bounded above by $\zeta$. That means, it suffices for $m$ to satisfy both Eqs.~\eqref{eqn35} and \eqref{eqn39}. Therefore we have
\begin{Lemma3}\label{lm3}
There are positive constants $c_4$, $c_5$ and $c_6$ independent of $d$, $\zeta$, $\eta$ and $\delta$, such that whenever
\begin{equation}\label{eqn40}
m > c_4 d^2 (c_5+\log \frac{1}{\eta})+ c_6 (\log\frac{1}{\zeta}+ \log\frac{1}{\delta}),
\end{equation}
the requirements of the protocol are satisfied.
\end{Lemma3}

Taking $m$ to be exactly of the order shown in on the right-hand-side of \eqref{eqn40}, then the requirements of the protocol are still satisfied.  The entanglement cost of the protocol can be expressed as
\begin{align}\label{eqn41}
& \log_2 \vert \TR_m \vert = O(m d^2) \notag\\
& = O(d_B^4 +d_B^4\log\frac{1}{\eta}+ d_B^2 \log \frac{1}{\zeta}+ d_B^2\log\frac{1}{\delta})
\end{align}
ebits. Here we have used the result in App.~\ref{appB} that $\vert \TR_m \vert$, the size of $\TR_m$, is equal to $2\times 6^{m d (d-1)/2}$. The following Theorem~\ref{thm2} shows that the three error parameters $\zeta, \eta, \delta$ can be combined into one parameter describing the deviation of the implemented quantum operation $\EC$ from the desired quantum operation $\EC_\TC$, the latter being defined as $\EC_\TC(\rho_{AB}):=\TC \rho_{AB} \TC^\dag$.

Before stating Theorem~\ref{thm2} we first introduce some notations. Suppose $T'$, $U'$ and $V'$ are the isometric extensions of $\EC_\TC$, $\EC_{\UC}$ and $\EC$, respectively:
\begin{align}\label{eqn42}
T' = \frac{1}{\sqrt{N}}\sum_{l=0}^{N-1}\TC\otimes \dyad{l}{0},\notag\\
U' = \frac{1}{\sqrt{N}}\sum_{l=0}^{N-1}\UC\otimes \dyad{l}{0},\notag\\
V' = \frac{1}{\sqrt{N}}\sum_{l=0}^{N-1}\UC_l \otimes \dyad{l}{0}.
\end{align}
where $\TC$, $\UC$ and $\UC_l$ act on the combined $AB$ system, and the operators $\dyad{l}{0}$ act on an ancillary system $R$. The set of kets $\{\ket{l}\}$ is (a part of) an orthonormal basis.

\begin{Theorem2}\label{thm2}
The quantity $\epsilon_0:=\|\EC_\TC-\EC\|_\diamond$ is related to $\zeta$, $\eta$ and $\delta$ in the following way:
\begin{equation}\label{eqn43}
\epsilon_0 \le 2\| T'-V'\|_\infty < 2 (\zeta + \sqrt{\eta^2 + 4\delta}).
\end{equation}
\end{Theorem2}

\begin{proof}
The first inequality is from Eq.~\eqref{eqn24}. In the following we prove the second inequality.
Define
\begin{equation}\label{eqn44}
D_i:= W_i - V_{k(i)}.
\end{equation}
Then from Eq.~\eqref{eqn37}, $\|D_i\|_\infty<\zeta$. From Eqs.~\eqref{eqn42}, \eqref{eqn34} and \eqref{eqn38} we have
\begin{align}\label{eqn45}
\| T'-U'\|_\infty &= \| \frac{1}{\sqrt{N}}\sum_{l=0}^{N-1}(\TC-\UC)\otimes \dyad{l}{0} \|_\infty \notag\\
&= \| \sum_{i=0}^{M-1}\dyad{i}{i}\otimes D_i \otimes \dyad{+}{0} \|_\infty,
\end{align}
where $\ket{+}:=\frac{1}{\sqrt{N}}\sum_{l=0}^{N-1}\ket{l}$ is a normalized pure state. Denote the operator inside the norm symbol in the last line by $K$, then
\begin{align}\label{eqn46}
\| K \|_\infty &= \sqrt{\lambda_0(K^\dag K)} \notag\\
& = \sqrt{\max_{0\le i\le M-1} \lambda_0(D_i^\dag D_i \otimes \dyad{0}{0})}\notag\\
& = \sqrt{\max_{0\le i\le M-1} \lambda_0(D_i^\dag D_i)} \notag\\
& < \sqrt{\max_{0\le i\le M-1} \zeta^2} = \zeta
\end{align}
where we have again used $\lambda_0(A)$ to denote the maximum eigenvalue of a Hermitian operator $A$, and again used the fact that the eigenvalues of a block diagonal matrix (where each block is a square matrix) are just the eigenvalues of the individual blocks. In deriving the last line we have used $\|D_i\|_\infty<\zeta$, and that $\|D_i\|_\infty^2$ is the maximum eigenvalue of $D_i^\dag D_i$.

And from Eq.~\eqref{eqn30}, $\| U'-V'\|_\infty \le \sqrt{\eta^2 + 4\delta}$ (note that this holds for the protocols shown in Fig.~\ref{fgr2} and Fig.~\ref{fgr3}, and still holds in the case of redundant terms, as remarked in the last paragraph in Sec.~\ref{sct3}). Hence
\begin{align}\label{eqn47}
\| T'-V'\|_\infty &\le \| T'-U'\|_\infty + \| U'-V'\|_\infty \notag\\
&< \zeta + \sqrt{\eta^2 + 4\delta}.
\end{align}
This proves the second inequality in Eq.~\eqref{eqn43}.
\end{proof}

We can always choose $\zeta, \eta, \delta$ to be small constants that are equal to each other, then $\log \frac{1}{\epsilon_0} \ge -\log{2} - \log (\delta+\sqrt{\delta^2+4\delta})\approx -\log{4} + \frac{1}{2} \log \frac{1}{\delta}$, hence $\log \frac{1}{\epsilon_0}$ is of at least the same order as $\log \frac{1}{\eta}$, when $\zeta=\eta=\delta$ is small. Therefore from \eqref{eqn41}, we get

\begin{Theorem3}\label{thm3}
When $\epsilon_0:=\|\EC_\TC-\EC\|_\diamond$ approaches $0$ and when $d_B$ approaches infinity, the asymptotic entanglement cost of the approximate fast controlled unitary protocol in this section is
\begin{equation}\label{eqn48}
O(d_B^4 \log \frac{1}{\epsilon_0})
\end{equation}
ebits.
\end{Theorem3}

Finally we remark on some practical issues in implementing the protocol. The construction of the local gates, such as $\hat R_k$, requires knowing the structure of the right quasigroup $Q$. Once the set $\TR_m$ is fixed, i.e. the operators $V_k$ in the approximate unitary representation of $Q$ are fixed, the structure of $Q$ is to be found out through the method in Appendix~\ref{appD}, and that depends a lot on how to find the best approximation for a given unitary using the unitaries in the set $\TR_m=\{V_k\}$. Each $V_k$ is a product of operators from the fixed generating set $\GC_d$. There is some work in the literature (e.g.~\cite{Nagy,Dawson}) on this topic of decomposing a given unitary into a product of gates from a fixed generating set; the paper by Nagy \cite{Nagy} includes a classical algorithm that runs faster than simply enumerating all the allowed products and picking the one closest to the target unitary. But our problem is somewhat different, as the operators $V_k$ do not include all possible products of unitaries from $\GC_d$. We leave open the question of whether the ideas in \cite{Nagy} are still useful in the current context.  There is some other work in the literature on the approximation of unitaries in $SU(d)$ with a finite quasigroup, see \cite{Gle02,Gle03}, but we have not applied the results of those papers to the current problem; on the other hand, the construction in this paper can be viewed as a scheme for approximation of the special unitary group $SU(d)$ by some right quasigroup.

\subsection{An example}
\label{sbct4.2}

Sometimes it might be possible to use some other constructions for the set $\{V_k\}$ than the one given in the previous subsection, to save entanglement cost or to achieve greater accuracy. This may happen when the controlled operators $W_k$ are simultaneously block-diagonal. The idea is to use the known constructions (e.g. those in Sec.~\ref{sbct4.1} or Sec.~\ref{sct2}) for each block, and then combine those. One specific example is as follows:\\

\textbf{Example 3.} Suppose the unitary to be implemented is on a $3\times 4$ dimensional space:
\begin{equation}\label{eqn49}
\TC=\sum_{k=0}^2 \dyad{k}{k}\otimes W_k=\sum_{k=0}^2 \dyad{k}{k}\otimes (W^{(1)}_k \oplus W^{(2)}_k),
\end{equation}
where $W^{(1)}_k$ are on a 2-dimensional subspace $\HC^{(1)}_B$, and $W^{(2)}_k$ are on a 2-dimensional subspace $\HC^{(2)}_B$ orthogonal to $\HC^{(1)}_B$.
$W^{(1)}_0=I_2, W^{(1)}_1=\exp(i \pi \sigma_x/3), W^{(1)}_2=\exp(i \pi \sigma_z/4)$,
and $W^{(2)}_0=W^{(2)}_1=I_2, W^{(2)}_2=\sigma_z$. ($I_2$ is the $2\times 2$ identity matrix; $\sigma_x$ and $\sigma_z$ are Pauli matrices.) We can choose a set $\{V^{(1)}_k:k\in Q_1\}$ on the subspace $\HC^{(1)}_B$ using the method in the previous subsection, where the size of the right quasigroup $Q_1$ is generally larger than 3, and $V^{(1)}_k$ with $k=0,1,2$ provide approximations to the corresponding $\{W^{(1)}_k\}$. Next, choose the set $\{Z_r:r=0,1\}=\{I_2,\sigma_z\}$ on $\HC^{(2)}_B$, which forms a representation of a cyclic group of order 2. Finally we take the direct sum of the operators as follows: $V'_{(k,r)}=V^{(1)}_k \oplus Z_r$. Then the set $\{V'_{(k,r)}:\,k\in Q_1,\,r=0,1\}$ is an approximate unitary representation of some right quasigroup, hence the protocol in the previous subsection can be applied to implement $\TC$ approximately. The size of $\{V'_{(k,r)}\}$ is just two times that of $\{V^{(1)}_k\}$, and is thus smaller than the size of $\{V_k:k\in Q\}$ that we would have obtained by choosing a ``generic'' approximate representation of a right quasigroup on the $4$-dimensional space $\HC_B$ (for the same degree of accuracy in approximation), hence this construction saves some entanglement resource.

\section{Comparing the fast protocols for controlled unitaries}\label{sct5}

\subsection{Adapting the nonlocal measurement protocol by Clark et al. to a fast unitary protocol}
\label{sbct5.1}

Clark {\it et al.} \cite{Clark10} introduced some instantaneous measurement protocols for bipartite observables. In this section we argue that the general protocol in Sec.~6 of that paper can be adapted to a fast protocol for implementing bipartite unitaries with similar entanglement cost.

In an instantaneous measurement protocol, each of the two parties performs a local measurement on his/her input system and part of the shared entangled state, and possibly some local ancillary system, and then they both send the measurement outcomes via classical channels to a third party, who calculates the final outcome of the nonlocal measurement from the classical input he receives.
It is not known if every instantaneous measurement protocol can be adapted to a fast nonlocal unitary protocol. For example, we do not know how to adapt some instantaneous measurement protocols for some special types of observables in \cite{Clark10} or in \cite{Groisman02} to fast nonlocal unitary protocols.

\begin{figure}[ht]
\begin{centering}
\includegraphics[bb=144 640 380 780, width=9.0 cm, clip]{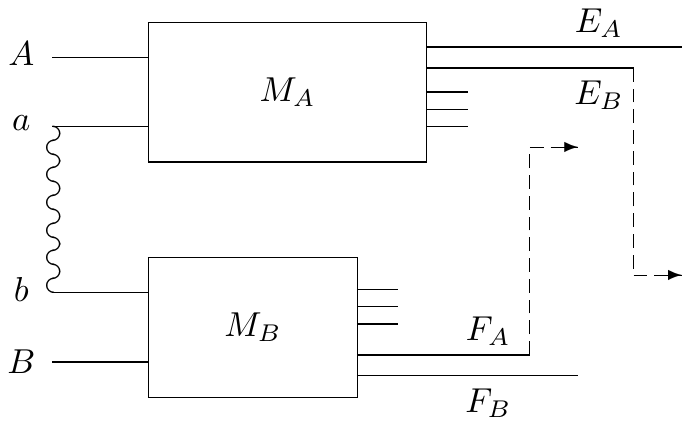}
\caption{Adapting the instantaneous measurement protocol in \cite{Clark10} to a fast unitary protocol. The figure only illustrates the simple case of a single Pauli rotation chain. The wavy line between $a$ and $b$ represents many pairs of maximally entangled qubits. Each party does some local operations (including measurements), as represented by $M_A$ and $M_B$ in the figure. Each party assumes that the true output state is in his/her own hands (illustrated as the $E_A E_B$ system or $F_A F_B$ system) at some termination point of the protocol. Alice teleports the $E_B$ subsystem to Bob, and at the same time Bob teleports the $F_A$ subsystem to Alice. Several steps after the teleportations are not shown in the figure: the parties communicate classically and find out where the actual output system is, and do local unitary corrections, then swap the respective part of output system to a fixed location.}
\label{fgr4}
\end{centering}
\end{figure}

However, the general instantaneous measurement protocol in Sec.~6 of \cite{Clark10} can indeed be adapted to a fast unitary protocol. Their protocol essentially does the following: first, perform a unitary (we will call it the \emph{target} unitary, since it is the unitary to be implemented in our adapted protocol) on the combined input system of Alice and Bob, with its output possessed by one of the parties only; and then Alice and Bob each does local measurements, and they send the measurement outcomes by classical means simultaneously to the other party.  Each party would have obtained enough information to figure out where the output of the target unitary was, and find out the outcome of the measurement on the output of that unitary, which is the outcome of their instantaneous measurement.   Our method of adapting the protocol is: remove the final measurement in the original protocol, and at the time when the ``target'' unitary is just implemented in the original protocol, do extra teleportations to distribute to each party one of the two subsystems of the output of the ``target unitary''.  These extra teleportations will be called ``delocalizing teleportations''. The details of this adapting process are discussed in the two paragraphs below.

We first describe the method for the simpler case of a single Pauli rotation chain, where a Pauli rotation chain is a sequence of local operations in the protocol in \cite{Clark10} which applies a Pauli rotation $R_{\bf j}(\theta) = \exp(-i \theta \sigma_{\bf j}/2) = \cos(\half\theta)\mathbf{I} - i\sin(\half\theta)\sigma_{\bf j}$, where $\sigma_{\bf j}$ is a Pauli string operator (tensor product of Pauli operators on qubits), and $\theta$ is a real number. If the target unitary is on two qubits, then it will be implementable by a single Pauli rotation chain if and only if the unitary is of the form $\cos(\half\theta) I_A \otimes I_B - i\sin(\half\theta) \sigma^A_z \otimes \sigma^B_z$ up to local unitaries, i.e. when it is of Schmidt rank 2. As illustrated in Fig~\ref{fgr4},  the output state of the ``target unitary'' is located in the $E_A E_B$ system or $F_A F_B$ system, and our method requires that Alice teleports $E_B$ to Bob while Bob teleports $F_A$ to Alice, just before doing the final measurement in the original protocol. Note that only one of the teleportations in the two directions will be actually sending the target state, because the target state will only be in the party that terminates ``earlier'' than the other party (``earlier'' is in the sense of the flow diagrams in \cite{Clark10}), but neither party knows if the other party had terminated and hence needs to act as if the target state is in his/her own hands. Then each party sends all the measurement outcomes (from all steps of the protocol so far) to the other party, and upon receipt of the classical messages each party figures out where the output system is. Without loss of generality, assume Alice terminated first. (For the case that Bob terminated first, the local subsystems on which the following operations are carried out will be different, hence our method is safe but a bit costly: Alice and Bob would need to do the operations corresponding to both cases.) By now the output system is already distributed to Alice and Bob, but still lacks two final local corrections, the \emph{first} one is only on Bob's side corresponding to Alice's measurement outcomes in the last teleportation step we added, and the \emph{second} one involves both parties and corresponds to the outcomes of Bob's last measurement (as a step of the original protocol). Then Alice and Bob do these local unitary corrections, with the ``first'' correction being done first, since it corresponds to a later measurement. Note that for the ``second'' local corrections to be local, the last measurement by Bob should be as if he were teleporting the two subsystems (belonging to Alice and Bob at the end) individually, and this is indeed the way in \cite{Clark10} that teleportation is done -- individually on each qubit. Of course, no final measurement on the output state is needed in our adapted protocol. In order for the output system to be at some fixed location no matter what the intermediate measurement outcomes are, a last step of the whole adapted protocol is that each of the two parties swaps his/her own part of the output system into a ``blank'' system at some fixed location.

Now let us discuss the more general case that more than one Pauli rotation chain is used in the instantaneous measurement protocol. The Pauli rotation chains are arranged in different levels, and each chain at a certain level is linked to many different chains in the next level. As a result, the final output state may be in many different places on the different chains on the final level. Now the final delocalizing teleportations must be performed on all the final-level Pauli-rotation chains, because each party cannot be sure on which of these chains the true output state is located, as the location generally depends on the other party's measurement results at various stages of the protocol. For each final-level chain, the added teleportation on each party need only be performed once, which is the same as in the single-chain case. The final steps are also similar as before: each party sends all measurement outcomes to the other party and figures out where the actual output state is, and does the appropriate local corrections, and swaps the respective part of the output system to a fixed location.

The extra entanglement cost resulting from the extra steps described above can be calculated as follows. Suppose the target unitary acts on a $d_A\times d_B$ space, then the part of the output system that needs to be teleported from Alice to Bob is of dimension $d_B$, and the part that needs to be teleported from Bob to Alice is of dimension $d_A$, and the last local swaps do not need entanglement, hence the extra entanglement cost is $\log_2 d_B$ ebits times the number of final-level Pauli rotation chains of Alice, plus $\log_2 d_A$ ebits times the number of final-level Pauli rotation chains of Bob. This quantity is not greater than the entanglement cost of the earlier parts of the protocol, hence the scaling behavior of the entanglement cost is the same as that of the instantaneous measurement protocol.

\subsection{Comparing the entanglement cost of various protocols}
\label{sbct5.2}

Now we compare the entanglement cost of the fast unitary protocols in (or adapted from) \cite{Buhrman,Beigi11,Clark10}, and our protocols in this paper, for implementing bipartite controlled unitaries. It should be noted that our protocols in this paper are specifically designed for controlled unitaries, while these other protocols are for general unitaries, but we apply them in the restricted case of controlled unitaries, and we will discuss the actual entanglement cost of these protocols for controlled unitaries, which is smaller than the cost for general unitaries for some of these protocols. Another point worth mentioning, since it will not be mentioned in the comparison of asymptotic entanglement cost in the following paragraphs, is that our controlled-group protocol performs well when the controlled operators form a representation of a small group.  For example, $\UC=\sum_{k=0}^{1} \dyad{k}{k} \otimes E_k$, with $E_1=I,\quad E_2=\mbox{diag}(1,e^{2\pi i/3})$ can be implemented by our protocol in Sec.~\ref{sct2} (or the controlled-cyclic-group protocol in Sec. II~A of \cite{nluf}) using only $\log_2 3$ ebits, while the protocols in \cite{Buhrman,Beigi11,Clark10} would need higher average-case (or worst-case) entanglement cost.

For general controlled unitaries, the entanglement cost of our approximate protocol in Sec.~\ref{sct4} is $O(d_B^4 \log \frac{1}{\epsilon_0})$ ebits. For the fast unitary protocol in Beigi and Konig \cite{Beigi11}, the error parameter $\epsilon$ in their Theorem~III.1 can be identified with our $\epsilon_0$, and that theorem gives an entanglement cost of $O(\frac{1}{\epsilon^2}d_B^8 \log d_B)$ ebits (assuming $d_A=d_B$) for general unitaries, and we have not found a specialized version of their protocol for controlled unitaries with improved entanglement cost. Hence our protocol has lower entanglement cost, by only comparing the two costs listed above.

Clark {\it et al.} \cite{Clark10} contains an instantaneous measurement protocol that can be adapted to a fast unitary protocol, as discussed in Sec.~\ref{sbct5.1}. This fast unitary protocol has a variable entanglement cost that depends on the measurement outcomes in the protocol. The average-case entanglement cost is exponential in $d_A d_B$, but independent of the error parameter $\epsilon_0$. Controlled unitaries have a simpler
decomposition in terms of Pauli rotations than general bipartite unitaries, hence they require less entanglement than general unitaries, but the average-case entanglement cost is still exponential in $d_B$. Taking into account the dependence on $\epsilon_0$ in the cost of our protocol, we see that only in the case of very small $\epsilon_0$ (doubly-exponentially small in $d_B$), the cost of our protocol is higher than that of the Clark {\it et al.} protocol.

The fast unitary protocol in Sec.~4 of Buhrman {\it et al.} \cite{Buhrman}, which is based on the instantaneous measurement protocol in \cite{Vaidman}, has an entanglement cost upper bounded by the exponential of a polynomial function of $d_A d_B$, while the dependence on the error probability $\epsilon$ is polynomial in $\frac{1}{\epsilon}$ (this $\epsilon$ can also be identified with $\epsilon_0$), which is unlike the cost of our protocol which is linear in $\log \frac{1}{\epsilon_0}$. We have not found a specialized version of Buhrman {\it et al.} protocol for controlled unitaries with better entanglement cost.

In summary, our protocols are quite efficient for general controlled unitaries, although these other protocols mentioned above might have improved versions specifically designed for controlled unitaries that have low entanglement cost. The entanglement cost also depends a lot on the specific form of the unitary. As a result, for a randomly chosen controlled unitary, the entanglement cost of our protocols may be less or more than the other protocols, depending on the form of the unitary and the required precision.

\section{Conclusions}\label{sct6}

In this paper, we have introduced two fast protocols for bipartite controlled unitaries. The first protocol can only implement the so-called ``controlled-group'' unitaries exactly, where the controlled operators form a subset of a projective representation of a finite group. The difference compared to the controlled unitary protocols in \cite{nluf} is that the group could be non-Abelian, and the representation could be projective. The second protocol approximately implements controlled unitaries using an algebraic structure called right quasigroup. This protocol could optionally use classical shared randomness as a nonlocal resource in additional to using entanglement. It is shown that the second protocol can implement \emph{any} controlled unitary approximately. The entanglement cost of this protocol for general controlled unitaries is compared with other fast unitary protocols in the literature. The cost of our protocol is small when the controlled operators form a representation of a small group, and the scaling behavior with the error parameter is not too bad.

A general open problem is to find the lower bound of entanglement cost to implement a given unitary in the fast way. It might be easier to first work on some special types of unitaries. Specifically, for the controlled-group unitaries $\UC=\sum_{k=0}^{M-1} \dyad{k}{k} \otimes V_k$, where the operators $V_k$ (exactly) form a subset of a projective representation of a finite group $G$, the smallest entanglement cost we know of is $\log_2 \vert G\vert$ ebits for general choices of the $V_k$, with two exceptions: when $\{V_k\}$ is a subset of a projective representation of a \emph{subgroup} of $G$, the subgroup could be used instead of $G$; or when the dimensions of $\HC_A$ and $\HC_B$ are small so that other protocols utilizing the idea of teleportation (e.g. \cite{Buhrman,Clark10,Beigi11}) might have lower cost.


For the protocol in Sec.~\ref{sct3}, there is an extension to the case of approximate \emph{projective} (as opposed to ordinary) unitary representations of right quasigroups, with phase factors defined in a way analogous to Eq.~\eqref{eqn3}, but we chose not to discuss that, as this extension might not change the entanglement cost in Theorem \ref{thm3} for general controlled unitaries. However, it might reduce the cost for some special classes of controlled unitaries.

Other possible directions for further investigation include: To find specific applications of our protocols in position-based quantum cryptography \cite{Kent,Kent06,Lau,Buhrman}; to adapt our protocols to instantaneous measurement protocols (the paper \cite{Beigi11} is an example where a fast unitary protocol and an instantaneous measurement protocol have similar structure); to generalize to fast protocols for implementing multipartite unitaries; to generalize to fast protocols for nonlocal non-unitary quantum operations.

Another open problem is: Are quasigroups or right/left quasigroups useful for the implementation of general nonlocal unitaries or quantum operations? An equivalent concept, Latin squares, has found many applications, e.g. in teleportation and dense coding schemes \cite{Werner}, in constructing unitaries with certain ``maximally-entangling'' property \cite{Clar05}, and in quantum error correction \cite{Aly}. The problem of finding sets of orthogonal Latin squares is in some cases related to finding sets of mutually unbiased bases, but the two problems might not be equivalent \cite{Haya05,Pate09,Pate10}. We hope the use of quasigroups (or similar algebraic structures) could help solve more different types of problems in quantum information theory.

\section{Acknowledgments}

The author thanks Robert Griffiths, Scott Cohen, Stephen Clark, Aram Harrow, Dan Stahlke, Joseph Fitzsimons, Christian Schaffner, and anonymous referees for helpful comments. This work was supported by the U.S. National Science Foundation through Grant No.~PHY-1068331 and by the National Research Foundation and Ministry of Education in Singapore.

\begin{appendix}

\section{The $\Lambda$ metric for finite sets of unitaries}\label{appA}

As a preparation for the next two appendices, in this appendix we review the definition of a quantity $\Lambda(\AC)$ introduced in Sec.~II of \cite{Harrow}, where $\AC$ is an arbitrary finite set of unitaries in $SU(d)$. It is a metric of how well products of unitaries drawn from the finite set $\AC$ or inverses of elements of $\AC$ approximate arbitrary elements of $SU(d)$, but in some sense it also describes how uniform the products of unitaries taken from $\AC$ or inverses of elements of $\AC$ is distributed in $SU(d)$. For the latter, see the last paragraph of this appendix and Appendix~\ref{appD} below.

Let $dg$ be the Haar measure on $SU(d)$ normalized so that $\int dg=1$.  Consider the Hilbert space $L^2(SU(d))$ with norm defined by the usual inner product $\langle \psi,\varphi\rangle\equiv\int\psi(g)^*\varphi(g)dg$.
The norm of a linear transformation on $L^2(SU(d))$ is given by
\begin{equation}\label{eqn50}
	|M|\equiv\sup\left\{\|Mf\|\big\arrowvert
	f\in L^2(SU(d)),\|f\|=1\right\}
\,.
\end{equation}
When $M$ is bounded and hermitian, the norm is simply the supremum of
its spectrum and as a result, $|M^n|=|M|^n$.

Define a representation $U\mapsto \tilde{U}$ of $SU(d)$ on
$L^2(SU(d))$ by
\begin{equation}\label{eqn51}
	\tilde{U}f(x)=f(U^{-1}x)
\,.
\end{equation}
Using the right invariance of the Haar measure, we see that
$\tilde{U}$ is unitary.  For any finite set $\AC \subset SU(d)$,
define the mixing operator $T(\AC)$ by
\begin{equation}\label{eqn52}
	T(\AC) = \frac{1}{2|\AC|}\sum_{A\in \AC} (\tilde{A} + \tilde{A}^{-1})
\,.
\end{equation}
All such $T$ are hermitian and have norm one.  We will often simply
write $T$ instead of $T(\AC)$.  These represent averaging the action
of the elements of $\AC$ and their inverses on a function; when the
function is a probability distribution on $SU(d)$ we can think of $T$
as multiplying by a random element of $\AC$ or its inverse.

We also define the following operator to be used in Appendix~\ref{appD}:
\begin{equation}\label{eqn53}
	T'(\AC) = \frac{1}{|\AC|}\sum_{A\in \AC} \tilde{A}
\,.
\end{equation}

Applying $T^n$ represents averaging over the action of products of length
$n$.  Denote the set of products of length $n$ made up of elements of
$\AC$ and their inverses by $W_n(\AC)$, or when the set $\AC$ is
understood, simply $W_n$. If two such products are equal to each other, we still count them as distinct ones, so that the whole set comprises $(2|\AC|)^n$ words. For any positive integer $n$, expanding $T^n$ gives
\begin{equation}\label{eqn54}
	T^n = \sum_{w\in W_n} \frac{\tilde{w}}{(2|\AC|)^n}
\,.
\end{equation}

We want to compare $T^n$ to the integral operator $P$, which is defined
as follows:
\begin{equation}\label{eqn55}
	Pf(h)=\int f(gh) dg = \int f(g) dg, \,\,\forall f\in L^2(SU(d)).
\end{equation}
Note that $P$ is the projection operator onto the set of constant
functions on $SU(d)$, and hence $P=P^\dag$ and $P^2=P$.
It is not hard to show that $TP=P=PT$ and consequently
\begin{equation}\label{eqn56}
	(T-P)^n=T^n-P
\,.
\end{equation}

The metric for comparing $T(\AC)$ to $P$ is given by
\begin{equation}\label{eqn57}
	\Lambda(\AC) \equiv |T(\AC)-P|
\,.
\end{equation}
From Eq.~\eqref{eqn56} and the hermiticity of $T$ and $P$, it
follows that
\begin{equation}\label{eqn58}
	\Lambda(\AC)^n = |T^n(\AC)-P|
\,.
\end{equation}
If one thinks of $T^n$ as a Riemann sum then $\Lambda$ serves as to
quantify how quickly $T^n$ converges to $P$.

Here we add some remarks which are not in \cite{Harrow}. From the definition of $P$, it is easy to see that $P$ could be thought of as the limit of the mixing operator formed by a large set of unitaries, where the set grows in size and approaches the Haar-uniform distribution in $SU(d)$. Combined with Eq.~\eqref{eqn58}, we see that whenever $\Lambda(\AC)<1$, the limit of a very long product of unitaries from $\AC$ or inverses of elements of $\AC$ would be indistinguishable from a random unitary chosen according to the Haar measure, hence the $\Lambda(\AC)$ measure describes how uniform the products of unitaries from $\AC$ or inverses of elements of $\AC$ is distributed in $SU(d)$.

\section{The set $R_m$}\label{appB}

Let us review the definition of the set of unitaries, $R_m$, in \cite{Harrow}, where $m$ is any positive integer.  Let
\begin{align}\label{eqn59}
V_1&=\frac{1}{\sqrt{5}}\left(\begin{array}{cc}1 & 2i \\ 2i & 1
\end{array}\right),\,
V_2=\frac{1}{\sqrt{5}}\left(\begin{array}{cc}1 & 2 \\ -2 & 1
\end{array}\right)\notag\\
\mbox{ and }V_3&=\frac{1}{\sqrt{5}}\left(\begin{array}{cc}1+2i & 0\\ 0&1-2i
\end{array}\right).
\end{align}
Then
$\lambda=\Lambda(\{V_1,V_2,V_3\})=\frac{\sqrt{5}}{3}<1$.
Let $\GC_2=\{V_1,V_2,V_3\}$. Let $I_k$ denote the $k\times k$ identity
matrix, then, for any $U\in SU(2)$ and $2\leq j\leq d$, define
$\beta^{(d)}_j(U)$ to be
\begin{equation}\label{eqn60}
	\beta^{(d)}_j(U)= \left(\begin{array}{ccc}
	I_{j-2} & 0 & 0 \\ 0 & U & 0 \\ 0 & 0 & I_{d-j} \end{array}\right)
	\in SU(d).
\end{equation}
\,
We will typically omit the $^{(d)}$ where it is understood.
For any $d>2$, define $\GC_d$ by
\begin{equation}\label{eqn61}
	\GC_d= \{\beta_j (V) \ | \ 1\leq j \leq (d-1), V\in\GC_2 \} \,.
\end{equation}

Let $R_m$ be the set of all products of the form
\begin{equation}\label{eqn62}
	\prod_{i=1}^{d-1}\prod_{j=i+1}^d \beta_j(G_j^i)
\end{equation}
such that the $G^i_j$ are selected from $W_m(\GC_2)$. (The $W_m$ notation is defined after Eq.~\eqref{eqn53}.)
For elements of $R_m$ that have different product forms but are actually identical, we treat them as distinct elements, hence the size of $R_m$ is exactly $6^{m d (d-1)/2}$. The set $\TR_m$ (see Sec.~\ref{sbct4.1}) consists of all elements of $R_m$ and their inverses, and again we do not check whether there are identical elements, hence the size of $\TR_m$ is exactly $2\times 6^{m d (d-1)/2}$.

The definition of $\TR_m$ and the definition of the $\Lambda$ metric immediately imply that $\Lambda(\TR_m)=\Lambda(R_m)$. Combined with the result of Eq.~(21) in \cite{Harrow}, this gives
\begin{equation}\label{eqn63}
	\Lambda(R_m) = \Lambda(\TR_m) \leq \frac{d(d-1)}{2}\lambda^m\,.
\end{equation}
We note here a small error in the middle expression in Eq.~(21) of \cite{Harrow}: the summation $w\in R_m$ should instead be $w\in \TR_m$, i.e. the inverses of elements of $R_m$ should be included in the summation as well, and accordingly the denominator should be changed from $\vert R_m\vert$ to $2\vert R_m\vert$. This error does not affect the correctness of the result of that equation when the middle expression is ignored.

\section{Approximating an arbitrary unitary in $SU(d)$ by an element of $\TR_m$}\label{appC}

Consider the set $\GC_2$ defined in Sec.~IV of \cite{Harrow}. According to the proof of Theorem 1 in \cite{Harrow}, any unitary in $SU(2)$ can be approximated with error bounded above by $\epsilon$ using the product of $m$ operators in $\GC_2$ whenever $m$ satisfies
\begin{equation}\label{eqn64}
m > \frac{2^2-1}{\log[1/\Lambda(\GC_2)]} (\log \frac{1}{\epsilon}+C_0)
\end{equation}
where $C_0$ is a positive constant. Since $\Lambda(\GC_2)=\frac{\sqrt{5}}{3}$, the inequality above can be written as
\begin{equation}\label{eqn65}
m > c_0+c_1\log \frac{1}{\epsilon},
\end{equation}
where $c_0$ and $c_1$ are positive constants.

From Lemma~3 in \cite{Harrow}, any unitary in $U\in SU(d)$ can be decomposed into a product of $d(d-1)/2$ unitaries, denoted by $U_k$, each of which is of the form in Eq.~(17) in \cite{Harrow}. This means each $U_k$ is block diagonal with only one $2\times 2$ block and all other blocks being $1\times 1$, where the $2\times 2$ block is a unitary in $SU(2)$, and all the $1\times 1$ blocks contain the element $1$. Thus we say $U_k$ is an extension of a unitary in $SU(2)$ of a form determined by the index $k$. Then from the previous paragraph, the $2\times 2$ block in $U_k$ can be approximated by a product of $m$ unitaries in $\GC_2$ for suitable values of $m$. Therefore for suitable values of $m$, $U_k$ can be approximated by $V_k$, a product of $m$ unitaries in $\GC_d$, each of which is an extension of a unitary in $\GC_2$ in the sense above. For any positive constant $\epsilon$, suppose $\|U_k-V_k\|_\infty<\epsilon$, then from Lemma~4 of \cite{Harrow}, $\|U_1 U_2 \cdots U_r - V_1 V_2 \cdots V_r\|_\infty<r\epsilon$, where $r=d(d-1)/2$, which means $\|U-V_1 V_2 \cdots V_r\|_\infty<\epsilon d(d-1)/2$. From Lemma 3 of \cite{Harrow} and the definition of $R_m$ in \cite{Harrow}, $V_1 V_2 \cdots V_r$ is in the set $R_m$, thus it is in $\TR_m$.  Let $\epsilon=\zeta/[d(d-1)/2]$. From the previous paragraph, there exist positive constants $c_0$ and $c_1$ such that whenever
\begin{equation}\label{eqn66}
m > c_0+c_1\log \frac{1}{\epsilon}=c_0+c_1\log \frac{d(d-1)}{2\zeta},
\end{equation}
$\|U_k-V_k\|_\infty<\epsilon$ holds, then $\|U-V_1 V_2 \cdots V_r\|_\infty<\epsilon d(d-1)/2=\zeta$. This proves Lemma~\ref{lm1}.

\section{Choosing a suitable right quasigroup for the protocol in Sec.~\ref{sct4}}\label{appD}

In this appendix, we prove Lemma~\ref{lm2}, showing that it is indeed possible to find a finite right quasigroup $Q$ such that the set $\{V_k\}=\TR_m$ with a suitable value of $m$ is an $(\eta,\delta)$-approximate unitary representation of $Q$. The operators $V_k$ all act on $\HC_B$, hence we denote $d:=d_B$ for simplicity.

A property of the set $\{V_k\}=\TR_m$ that will be exploited here is that its elements are almost uniformly distributed in the $SU(d)$ group for sufficiently large values of $m$ [``almost uniformly'' is used in a non-technical sense here; technically it can be expressed using the last two lines of Eq.~\eqref{eqn70} below].  Our aim is to find the structure of a right quasigroup $(Q,*)$ such that $\{V_k\}$ form an $(\eta,\delta)$-approximate unitary representation of $Q$ in the sense of Definition \ref{defi1}. For each $k$, we need to find at least $N(1-\delta)$ distinct values of $j$ such that $\| V_{l(j,k)} V_k - V_j\|_\infty <\eta$, where $N$ is the size of the set $\{V_k\}$ and also the size of $Q$. Since $l(j,k)*k=j$, from the definition of a right quasigroup, $j=l*k$ is unique for fixed $l$ and $k$, and $l$ is uniquely determined by $j$ and $k$. That means, for each fixed $k$, we need to find a bijection between $j$ and $l$ (one-to-one correspondence between $j$ and $l$) suitable for the ``approximate unitary representation'', where the possible values of $j$ and $l$ are between $0$ and $N-1$. Such bijections are usually different for different values of $k$, although this is not a requirement.  Such bijection is also called a bipartite perfect matching between the two sets of indices (viewed as vertices in a bipartite graph).

Such a perfect bipartite matching consists of two parts: at least $N(1-\delta)$ edges for unitaries $V_j$ and $V_l$ satisfying $\| V_l V_k - V_j\|_\infty <\eta$ ($k$ is fixed here), and the remaining edges in the matching do not have to satisfy this inequality. Hence the problem boils down to finding an imperfect matching containing at least $N(1-\delta)$ edges in a bipartite graph $G_k$, where the vertices of $G_k$ correspond to the unitaries $V_j$ and $V_l$ ($j,l\in \{0,1,\cdots,N-1\}$), and there is an edge between the vertices for $V_j$ and $V_l$ iff $\| V_l V_k - V_j\|_\infty <\eta$.

We will need to cite a theorem stated in \cite{Frederic} on bipartite matching, which is a generalization of Hall's theorem for perfect bipartite matching.

\begin{Theorem4}\label{thm4}
Let $G=((A,B),E)$ be a bipartite graph and $t\in \mathbb{N}$. $G$ has a matching of size $t$ if and only if $\vert N(S)\vert\ge \vert S\vert-\vert A\vert+t$  for all $S\subseteq A$.
\end{Theorem4}

In the theorem above, $A$ and $B$ are the two parts of the graph, and $N(S)$ is the \emph{neighborhood} of $S$, i.e. the set of vertices in $G\backslash S$ adjacent to at least one vertex of $S$. For a simple proof, see \cite{Frederic}.

Now we proceed with the proof of the main result of this appendix. We will use the $\Lambda$ metric defined for a finite set of unitaries. See Sec.~II of \cite{Harrow} or Appendix~\ref{appA} for the definition.

We will also need a fact about the geometry of $SU(d)$, which was noted in Sec.~III of \cite{Harrow}.  For any $d$ and $r_0$, if $V(r)$ is the Haar measure of a ball of radius $r$ in $SU(d)$, then there
exist constants $k_1$ and $k_2$ such that
\begin{equation}\label{eqn67}
	k_1 r^{d^2-1}<V(r)<k_2 r^{d^2-1}\,.
\end{equation}
for all $r\in(0,r_0)$.

Define a function $\chi\in L^2(SU(d))$ by
\begin{equation}\label{eqn68}
	\chi(g)=\left\{
    \begin{array}{ll}
      1/\sqrt{V}, & \mbox{for }|g-I|<\eta/2,\\
	  0, & \mbox{otherwise.}
    \end{array}\right.
\end{equation}
where $V$ is the Haar measure of the ball around the
identity of radius $\eta/2$, hence the function $\chi$ is normalized.
From Eq.~\eqref{eqn67} we have that $V>k_1(\eta/2)^{d^2-1}$. We will encounter other functions
similar to $\chi$ in that they are constant on a ball and zero elsewhere; these functions will be referred to as \emph{ball-functions}.

For any fixed $k$, Let $\TR^{(k)}_m:=\{U\cdot V_k \big\arrowvert U\in \TR_m\}$.  Let $T'_{1m}=T'(\TR_m)$ and $T'_{2m}=T'(\TR^{(k)}_m)$ using the definition in Eq.~\eqref{eqn53}.
We have $|T'_{2m}-P|=|T'_{1m}-P|$, which can be proved as follows:
\begin{align}\label{eqn69}
|T'_{2m}-P| &= \sup\left\{\|(T'_{2m}-P)f\| \big\arrowvert \|f\|=1\right\} \notag\\
&= \sup\left\{\|(T'_{1m}-P)({\tilde V}_k) f\| \big\arrowvert \|f\|=1\right\}\notag\\
&= \sup\left\{\|(T'_{1m}-P) ({\tilde V}_k f)\| \big\arrowvert \|f\|=1\right\}\notag\\
&= \sup\left\{\|(T'_{1m}-P) h)\| \big\arrowvert \|h\|=1\right\}\notag\\
&= |T'_{1m}-P|\,,
\end{align}
where we have abbreviated the requirement that $f,h\in L^2(SU(d))$. ${\tilde V}_k$ is an operator on $L^2(SU(d))$ defined by ${\tilde V}_k f(x)=f(V_k^{-1}x)$. In deriving the second line we have used $P {\tilde V}_k=P$ [see the paragraph after Eq.~\eqref{eqn55}]; and $T'_{2m}=T'_{1m} {\tilde V}_k$ follows from the definition of $R^{(k)}_m$.
In deriving the fourth line we used the replacement $h:={\tilde V}_k f$ and used the fact that ${\tilde V}_k$ is unitary on the space $L^2(SU(d)))$ [see the remark after Eq.~\eqref{eqn51}] which implies $\|h\|=\|f\|$.

According to Eq.~\eqref{eqn50},
\begin{align}\label{eqn70}
\| (T'_{1m}-T'_{2m})\chi \| &\le |T'_{1m}-T'_{2m}| \notag\\
&\le  |T'_{1m}-P| + |T'_{2m}-P| \notag\\
&= 2 |T'_{1m}-P| \notag\\
&= 2 \Lambda(\TR_m) \notag\\
&\le d(d-1)\lambda^m,
\end{align}
where in deriving the second line we have used the triangle inequality of the norm of Eq.~\eqref{eqn50}, which is a consequence of the triangle inequality of the $L^2$-norm of functions in $L^2(SU(d))$. And in deriving the last line we have used Eq.~(21) of \cite{Harrow} and also $\Lambda(\TR_m)=\Lambda(R_m)$ which holds by definition.

Note that $T'_{1m}\chi$ is a function which is an equal-weighted linear combination of many ball-functions with the radius of the balls being $\eta/2$; the locations of the centers of the balls are determined by elements of $\TR_m$. The function $T'_{2m}\chi$ has the same properties except that the centers of balls are determined by elements of $\TR^{(k)}_m$.

We are only going to prove that some values of $m$ are sufficient for there to be a desired matching, but will not study all possible values of $m$ that guarantee the existence of a desired matching, since the latter is unnecessary for proving Theorem~\ref{thm3}. The proof is by contradiction. Assume the maximum possible matching has size less than $N(1-\delta)$, then according to Theorem~\ref{thm4}, there must exist a set of indices $S_0$ in $A=\{0,1,\cdots,N-1\}$ (corresponding to a subset of $\TR_m$) such that
$|N(S_0)|<|S_0|-|A|+N(1-\delta)$, where $|A|=N$; this can be equivalently written as
\begin{equation}\label{eqn71}
|S_0|-|N(S_0)|>N\delta.
\end{equation}
The square of the left side of Eq.~\eqref{eqn70},
\begin{equation}\label{eqn72}
\| (T'_{1m}-T'_{2m})\chi \|^2 \equiv \int [\frac{1}{N}(\sum_j {\tilde V}_j-\sum_{j'} \widetilde{V_{j'} V_k})\chi(g)]^2 dg
\end{equation}
is an integral of a square function. $\widetilde{V_{j'} V_k}$ is defined by $\widetilde{V_{j'} V_k}f(x)=f[(V_{j'} V_k)^{-1}x]$, $\forall f\in L^2(SU(d))$. To study this integral it is helpful to first consider the integral of $(T'_{1m}-T'_{2m})\chi(g) = \frac{1}{N}(\sum_j {\tilde V}_j-\sum_{j'} \widetilde{V_{j'} V_k})\chi(g)$. Let the set $R\subset SU(d)$ be the set of points that are at most $\eta/2$ away from any of ball-centers that correspond to $V_j$ with $j\in S_0$. Denote the Haar measure of this set $R$ as $V(R)$. We have
\begin{equation}\label{eqn73}
\int_{R} (T'_{1m}-T'_{2m})\chi(g) dg > \sqrt{V}\delta,
\end{equation}
which can be shown using the following argument: the integrand is the sum of ball-functions of the form $\frac{1}{N} {\tilde V}_j\chi(g)$ or $-\frac{1}{N} \widetilde{V_{j'} V_k}\chi(g)$ where the balls are of radius $\eta/2$ and centered at $V_j$ or $V_{j'} V_k$. A negative-valued ball-function of the type $-\frac{1}{N} \widetilde{V_{j'} V_k}\chi(g)$ would have some overlap with $R$ if and only if it overlaps with a positive-valued ball-function of the type $\frac{1}{N} {\tilde V}_j\chi(g)$ with $j\in S_0$, and the latter means the centers of the two balls have distance less than $\eta$, i.e. $|V_j-V_{j'} V_k|<\eta$, hence there is an edge in the bipartite graph $G_k$ connecting the two vertices corresponding to these two ball-centers, therefore the center of any ball for the negative-valued ball-function that have overlap with $R$ correspond to a vertex in $N(S_0)$. Since the positive-valued balls with indices $j\in S_0$ appear completely inside $R$ while the negative-valued balls are generally not completely inside $R$, the integral $\int_{R} (T'_{1m}-T'_{2m})\chi(g) dg=\int_R \frac{1}{N}[\sum_{j\in S_0} {\tilde V}_j-\sum_{w(j',k)\in N(S_0)} \widetilde{V_{j'} V_k}]\chi(g) dg$ is not less than the integral of the right-hand-side extended from $R$ to the whole $SU(d)$ space [where $w(j,k)$ is the label for the element $V_{j'} V_k$ in $\TR'_m$], and  the latter integral is equal to $\frac{1}{N}\sqrt{V}(|S_0|-|N(S_0)|)>\frac{1}{N}\sqrt{V}N\delta=\sqrt{V}\delta$,
where we have used that the absolute value of the integral of each ball-function is $\frac{1}{N}V/\sqrt{V}=\frac{1}{N}\sqrt{V}$, and used the inequality \eqref{eqn71}. Hence the inequality \eqref{eqn73} holds. Therefore
\begin{align}\label{eqn74}
\| (T'_{1m}-T'_{2m})\chi \| &\ge {\left( \int_R [(T'_{1m}-T'_{2m})\chi(g)]^2 dg\right)}^{1/2} \notag\\
&\ge {\left(\int_{R} (T'_{1m}-T'_{2m})\chi(g) dg \right)} / (\int_R dg)^{1/2} \notag\\
&> \sqrt{V}\delta / [V(R)]^{1/2} \notag\\
&\ge \delta \sqrt{V} \notag\\
&> \sqrt{k_1} \delta (\eta/2)^{(d^2-1)/2},
\end{align}
where in deriving the second line we have used the Cauchy-Schwarz inequality for integrals of continuous real functions:
\begin{equation}\label{eqn75}
[\int f(g)^2 dg] [\int h(g)^2 dg] \ge [\int f(g)h(g) dg]^2,
\end{equation}
and for our problem we have assumed $f(g):=(T'_{1m}-T'_{2m})\chi(g)$ and $h(g):=1$.
In deriving the second-last line of Eq.~\eqref{eqn74} we have used $V(R)\le 1$.

From Eqs.~\eqref{eqn74} and \eqref{eqn70},
\begin{equation}\label{eqn76}
\sqrt{k_1} \delta (\eta/2)^{(d^2-1)/2} < d(d-1)\lambda^m,
\end{equation}
where $\lambda=\frac{\sqrt{5}}{3}$. Taking the logarithm and multiply by $(-1)$ on both sides, we obtain
\begin{equation}\label{eqn77}
c + \log \frac{1}{\delta} + \frac{d^2-1}{2} \log \frac{2}{\eta} > -\log {d(d-1)}+ m \log\frac{1}{\lambda}
\end{equation}
for some constant $c$. Hence there exists a positive constant $c_3$ such that whenever
\begin{equation}\label{eqn78}
m > c_3 \cdot (d^2 \log\frac{2}{\eta}+\log\frac{1}{\delta}),
\end{equation}
the previous inequality \eqref{eqn77} would not hold, and hence the assumption that the maximum possible matching has size less than $N(1-\delta)$ is not true, therefore there is a matching of size at least $N(1-\delta)$. As the above argument works for all $k$, there exists a desired matching for any $k$ when Eq.~\eqref{eqn78} holds, hence the set $\TR_m$ is an $(\eta,\delta)$-approximate unitary representation of the right quasigroup $(Q,*)$, where the multiplication rule of the right quasigroup depends on the details of $\TR_m$.

\end{appendix}

\end{document}